\documentclass[aps,prl,twocolumn,showpacs,groupedaddress]{revtex4}
\usepackage{amssymb}
\usepackage{amsmath}
\usepackage{graphicx}
\usepackage{hyperref}
\usepackage{epstopdf}

\newcommand{\vb}{\mathbf{b}}

\newcommand{\vs}{\mathbf{s}}

\begin{document}

\title{Quantum strings in quantum spin ice}

\author{Yuan Wan}
\author{Oleg Tchernyshyov}
\affiliation{Department of Physics and Astronomy, Johns Hopkins University, Baltimore, MD 21218}

\date{\today}


\begin{abstract}
We study quantum spin ice in an external magnetic field applied along a $\langle 100 \rangle$ direction. When quantum spin fluctuations are weak, elementary excitations are quantum strings with monopoles at their ends manifested as multiple spin-wave branches in the dynamical structure factor. Strong quantum fluctuations make the string tension negative and give rise to the deconfinement of monopoles. We discuss our results in the light of recent neutron scattering experiments in $\mathrm{Yb_2Ti_2O_7}$.
\end{abstract}

\maketitle

The quest for novel quantum phases and elementary excitations is one of the central themes in condensed-matter physics. The notion of an elementary excitation is conventionally associated with a point-like object, as the term \emph{quasiparticle} implies. A natural question is whether elementary excitations in quantum materials could resemble \emph{strings}, rather than particles. String excitations were recently found in spin ice $\mathrm{Dy_2Ti_2O_7}$ \cite{Jaubert, DyTiO}, a frustrated ferromagnet with fractionalized excitations known as magnetic monopoles \cite{Ryzhkin, Castelnovo}. In an applied magnetic field, excitations are strings of misaligned spins connecting two monopoles of opposite charge. 

Conventional spin ice is a classical magnet with Ising spins \cite{Gingras}. Therefore, magnetic monopoles and strings in it are classical objects whose dynamics are due to thermal fluctuations. In this letter, we propose that string excitations with inherent quantum dynamics may exist in quantum spin ice, a new family of spin-ice materials exemplified by $\mathrm{Tb_2Ti_2O_7}$ and $\mathrm{Yb_2Ti_2O_7}$ \cite{TbTiO}\cite{Thompson2}. In these compounds, spins exhibit substantial quantum fluctuations. We demonstrate that, in a certain regime of coupling constants, elementary excitations of quantum spin ice are strings with quantum dynamics. The calculated dynamical structure factor $S(\omega,\mathbf k)$ reveals multiple branches of excitations that correspond, loosely speaking, to strings of different lengths. As the applied field increases, these branches gradually separate and the lowest one evolves into a magnon. We connect these findings to recent experiments on neutron scattering in $\mathrm{Yb_2Ti_2O_7}$ \cite{YbTiO, YbTiO-FM}.

\begin{figure}
\includegraphics[width=0.4\textwidth]{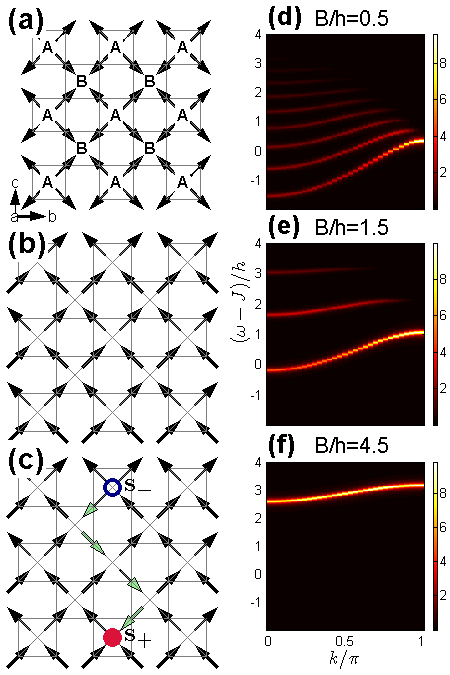}
\caption{(a) The checkerboard lattice. A and B denote two symmetrically inequivalent planar tetrahedra, and arrows the local $\hat{\mathbf z}_i$ directions.  (b) The fully-polarized state when the field is applied in the $c$ direction. Arrows denote the spin orientations. (c) A string of flipped spins (light green) binding a $Q=+1$ monopole (red solid circle) and a $Q=-1$ one (open blue circle). (d, e, f) $-\textrm{Im}S^{aa}(\omega,\mathbf{k})$ for $k_b =0$. $B/h=0.5$, 1.5, and 4.5 respectively.}
\label{fig:ice-2d}
\end{figure}

We begin with a toy model of quantum spin ice on the two-dimensional checkerboard lattice, Fig. \ref{fig:ice-2d}. The point of departure is classical spin ice, in which spins have projections $S_i^z = \pm 1/2$ on local directions $\hat{\mathbf z}_i$ shown in Fig. \ref{fig:ice-2d}a. Magnetic charge on a crossed plaquette (planar tetrahedron) is defined as $Q_\boxtimes = -\epsilon_{\boxtimes}\sum_{i\in\boxtimes}S_i^z$, with $\epsilon_{\boxtimes} = \pm 1$ for sublattice A (B). The ground states of the classical spin-ice Hamiltonian,
\begin{equation}
H_0 = \sum_\boxtimes \sum_{\langle ij\rangle \in \boxtimes} J S_i^z S_j^z 
= \sum_{\boxtimes} JQ_\boxtimes^2/2 + \mathrm{const},
\label{eq:H-square-ice-classical}
\end{equation} 
obey the Bernal-Fowler rule, $Q_\boxtimes=0$, on every tetrahedron \cite{Gingras}. Next we apply a weak magnetic field in the $ac$ plane. In the local frames, the perturbation reads
\begin{equation}
H_1 = - \sum_{i}(h S_i^x + B \eta_i S_i^z). 
\label{eq:H-square-ice-quantum}
\end{equation}
Here we chose the local $y$-axes to be orthogonal to the field and introduced cosines $\eta_i \equiv \hat{\mathbf c} \cdot \hat{\mathbf z}_i = (-1)^{c_i}/\sqrt{2}$. The Zeeman term (\ref{eq:H-square-ice-quantum}) has two effects. Its longitudinal component $B$ breaks the degeneracy of ice states and favors a fully magnetized state, Fig.~\ref{fig:ice-2d}b. The transverse component $h$ induces quantum fluctuations of spins. We treat $B$ and $h$ as independent parameters in the toy model.

Flipping a single spin in the fully magnetized state creates two monopoles with $Q=\pm1$, which can be pulled further apart. The process creates a string of spins aligned against the field and connecting the monopoles, Fig.~\ref{fig:ice-2d}c. For $h=0$, the energy of a string with $n$ segments is $J+Bn/\sqrt{2}$. For weak fields, the Hilbert space thus separates into near-degenerate subspaces with a fixed number of strings. The transverse part of the Zeeman term (\ref{eq:H-square-ice-quantum}) mixes states in the same subspace through quantum tunneling, inducing quantum motion of strings. We use degenerate perturbation theory in the subspace with a single string to construct an effective theory of its quantum dynamics.

The shape of a string is specified by its segments $\{\mathbf{s}_1, \mathbf{s}_2\ldots \mathbf{s}_{n}\}$, or $\{\mathbf s_i\}$ for short, which take on the values $\mathbf r \equiv(0,1,1)$ and $\mathbf l \equiv(0,-1,1)$ in the $abc$-frame. The string thus propagates upwards in Fig.~\ref{fig:ice-2d}c from the $Q=+1$ monopole at $\mathbf s_+$ to the $Q=-1$ monopole at $\mathbf s_-$. Because of the constraint $\mathbf s_- - \mathbf s_+ = \sum_{i=1}^n \mathbf s_i$, the state of a string is fully specified by its shape and location of one of the ends, $|\mathbf s_+, \{\mathbf s\}\rangle$. We introduce a hybrid basis with fixed shape $\{\mathbf{s}_i\}$, $c$-coordinate of the monopole $c_{+}$, and  the $b$-component of the total momentum $k_b$:
\begin{equation}
|k_b, c_{+}, \{\mathbf{s}_i\}\rangle 
	= \sum_{{b}_+}e^{ik_b(b_+ + b_-)/2}
	|b_+, c_+, \{\mathbf{s}_i\}\rangle.
\end{equation}

\begin{figure}
\includegraphics[width=0.4\textwidth]{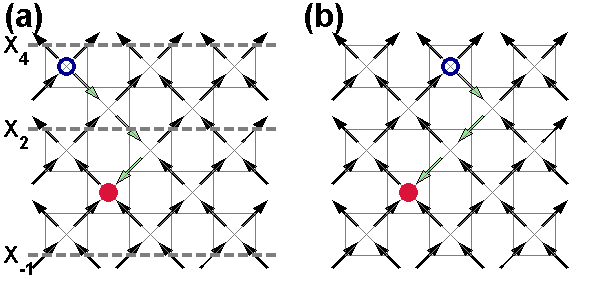}
\caption{Definition of $X_c$ operators. The operator $X_{2}$ changes the orientation of the string segment with $c=2$ from $\mathbf{l}$ to $\mathbf{r}$ in state (a) and results in a new state (b). The operators $X_{-1,5}$ fall outside the range of the string and act trivially on (a).}\label{swap}
\end{figure}

To the first order in $h$, the motion of a string involves removing or adding  a segment at one of the ends, with an effective Hamiltonian 
\begin{widetext}
\begin{eqnarray}
H_\mathrm{eff}|c_{+},\{\mathbf{s}_1\ldots \mathbf{s}_{n}\}\rangle 
&=& (J + nB/\sqrt{2})|c_{+},\{\mathbf{s}_1\ldots\mathbf{s}_{n}\}\rangle 
-(h/2)e^{ik_bb_{n}/2}|c_{+},\{\mathbf{s}_1\ldots\mathbf{s}_{n-1}\}\rangle
-(h/2)e^{-ik_bb_1/2}|c_{+}+1,\{\mathbf{s}_2\ldots\mathbf{s}_{n}\}\rangle
\nonumber\\	
&&  - \ (h/2) \sum_{\mathbf{s}_{n+1}}e^{-ik_bb_{n+1}/2}|c_{+},\{\mathbf{s}_1\ldots\mathbf{s}_{n+1}\}\rangle
- \ (h/2) \sum_{\mathbf{s}_{0}}e^{ik_bb_0/2}|c_{+}-1,\{\mathbf{s}_0\ldots\mathbf{s}_{n}\}\rangle.
\label{eq:Heff-square}
\end{eqnarray}
\end{widetext}
Here $b_{i}$ stands for the b-component of the vector $\mathbf{s}_i$. We have omitted the momentum index to simplify the notation.

When $k_b = 0$, diagonalization of $H_\mathrm{eff}$ is simplified by the presence of multiple reflection symmetries. Define the parity operator $X_c$ that switches between the $\mathbf{l}$ and $\mathbf{r}$ orientations of the segment with coordinate $c$ and keeps all other segment variables $\mathbf s_i$ intact (Fig.\ref{swap}), e.g.
\begin{equation}
X_c|c_+,\dots\mathbf{s}_{c-c_+},\mathbf{l}\dots\rangle
=|c_+,\dots\mathbf{s}_{c-c_+},\mathbf{r}\dots\rangle.
\end{equation}
When $X_c$ falls outside the range of the string, $c_+ < c < c_-$, it acts on the vacuum state, which is symmetric, so we set $X_c|c_+,\{\mathbf{s}\}\rangle = +|c_+,\{\mathbf{s}\}\rangle$ in this case. It can be seen that $X^2_c=1$ and $[X_c,X_{c'}]=0$. Although $X_c$ does not preserve the coordinate of the other end of the string $\mathbf s_-$, at $k_b = 0$ its horizontal displacement makes no difference; therefore, $[X_c,H_\mathrm{eff}]=0$. Thus, all $k_b=0$ eigenstates of $H_\mathrm{eff}$ can be classified by their parities under $\{X_c\}$ and $H_\mathrm{eff}$ becomes block-diagonal.The most important states have all even parities, $X_c = +1$. An all-even state of a string of length $n$ and longitudinal momentum $k_c$ is
\begin{equation}
|k_c, n\rangle 
	= 2^{-n/2}\sum_{c_+}\sum_{\mathbf{s_1}\dots\mathbf{s_n}}
	e^{ik_c(c_+ + c_{-})/2}|c_+,\{\mathbf s_1...\mathbf s_n\}\rangle.
\end{equation}
For them, the Hamiltonian (\ref{eq:Heff-square}) simplifies,
\begin{equation}
H_\mathrm{eff}|n\rangle = \left(J + \frac{nB}{\sqrt{2}}\right)|n\rangle
- \sqrt{2}h\cos{\frac{k_c}{2}} \sum_{m=n\pm{}1}|m \rangle.
\end{equation}
The above is equivalent to the one-dimensional problem of a particle on a lattice subject to a constant force $-B/\sqrt{2}$ and a hard wall at $n=0$. For $B \ll h$, we use the continuum approximation to find the spectrum:
\begin{equation}
E_{j}(k_c) = J - 2\sqrt{2}\left|h\cos{\frac{k_c}{2}}\right|
	+\lambda_j \left|\frac{\sqrt{2}}{2}B^2h\cos{\frac{k_c}{2}}\right|^{1/3}.
\end{equation}
Here $\lambda_j$ are roots of the Airy function. When $B \gg h$, the lowest eigenstate is a single misaligned spin with the dispersion
\begin{equation}
E_1(k_c) = J + \frac{B}{\sqrt{2}}\ 
	-\frac{\sqrt{2} h^2}{B}(1 + \cos{k_c}).
\end{equation}
Likewise, $H_\mathrm{eff}$ can be diagonalized in odd-parity sectors \cite{SP}. 

Strings can be directly observed in neutron scattering experiments. A scattered neutron flips a spin in the fully-polarized background, creating a string of length 1. The intensity of scattering is proportional to the overlap between a length-1 string and a string eigenstate of $H_\mathrm{eff}$ squared. Fig.~\ref{fig:ice-2d} shows the dynamical structure factor $-\textrm{Im}S^{aa}(\omega,\mathbf{k})$ at several values of $B/h$ for $k_b = 0$. For this direction of $\mathbf k$, the spectral weight comes solely from states with all-even parities, $X_m = +1$. For $B \lesssim h$, the spectrum consists of overlapping bands, whereas for $B \gg h$ the bands separate and the spectrum becomes dominated by the shortest string consisting of a single flipped spin, in essence a magnon.

For general $\mathbf{k}$, we used the Lanczos method to calculate the spectrum numerically and found similar behavior. Parities $X_c$ are no longer good quantum numbers; therefore, more bands appear in the spectrum.

\begin{figure}
\includegraphics[width=0.4\textwidth]{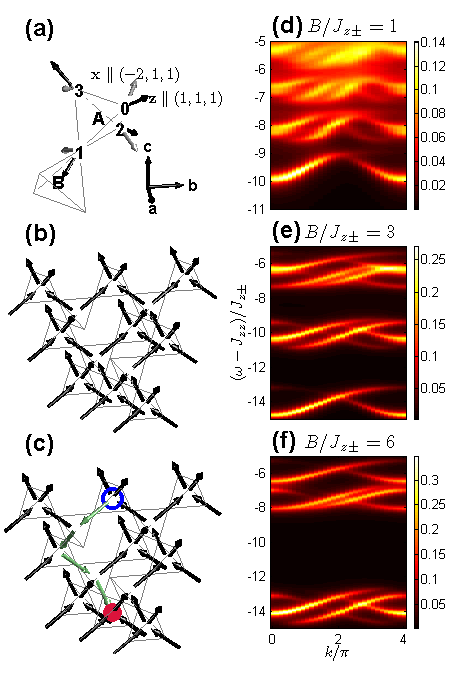}
\caption{(a) A and B denote two inequivalent tetrahedra in the pyrochlore lattice and $0\sim{}3$ four sublattices. The gray and black arrows show the local $\hat{\mathbf x}$ and $\hat{\mathbf z}$ directions. The $abc$ vectors specify the local frame for one sublattice. (b) The fully-polarized state when the field is applied in the $c$ direction. Arrows show the spin orientations. (c) A string of flipped spins (light green) binding a $Q=+1$ monopole (red solid circle) and a $Q=-1$ one (blue open circle) (d,e,f) The neutron scattering spectra for the momentum transfer $\mathbf{k}\parallel{}\mathbf{B}$. $B/J_{z\pm}=1$,3, and 6 respectively.}
\label{fig:ice-3d}
\end{figure}

The case of three-dimensional quantum spin ice, with $S=1/2$ spins on the pyrochlore lattice, proceeds along similar lines. The most general exchange Hamiltonian is written in local axes (Fig.~\ref{fig:ice-3d}a) as \cite{Curnoe}
\begin{eqnarray}
H_\mathrm{pyro}&=&\sum_{\langle{}ij\rangle}J_{zz}S^z_iS^z_j-J_{z\pm}[S^z_i(\zeta_{ij}S_j^++\zeta^\ast_{ij}S_j^-)+(i\leftrightarrow{}j)]\nonumber\\
&&-J_{\pm}(S_i^+S_j^-+h.c.)-J_{\pm\pm}(\zeta^\ast_{ij}S^+_iS^+_j+h.c.)
\end{eqnarray}
Here $\zeta_{ij}=\zeta_{ji}$ are phase factors, and $i$ and $j$ labeling spin sublattices 0 to 3.  Specifically, $\zeta_{01}=\zeta_{23}=-1$, $\zeta_{02}=\zeta_{13}=\exp(i\pi/3)$, $\zeta_{03}=\zeta_{12}=\exp(-i\pi/3)$, and $\zeta_{ii}=0$. The $J_{zz}$ term describes classical spin ice, whereas the three remaining terms create quantum fluctuations. 

A magnetic field applied in the $[001]$ direction adds the Zeeman term $-B\sum_{i}\alpha_i S^x_i+\beta_i S^y_i+\gamma_i S^z_i$, with the cosines 
\begin{eqnarray}
\alpha_{0,3} = - \alpha_{1,2} = \frac{g_{xy}}{g_{z}\sqrt{6}}, &&
\quad \beta_{0,3} = - \beta_{1,2} = \frac{g_{xy}}{g_{z}\sqrt{2}}, 
\nonumber\\
\gamma_{0,3} = - \gamma_{1,2} = \frac{1}{\sqrt{3}}, &&
\end{eqnarray}
where $g_{xy}$ and $g_{z}$ are the principal components of the $g$-tensor. In what follows we assume that the spin-ice term $J_{zz}$ dominates and treat the rest of the terms as perturbations. The $z$ Zeeman term favors the fully-magnetized state (Fig.~\ref{fig:ice-3d}b). Excitations are open strings connecting a pair of monopoles with $Q=\pm1$. Magnetic charge is defined as usual, $Q_{\boxtimes} \equiv -\epsilon_{\boxtimes}\sum_{i\in\boxtimes}S^c_i$, where $\boxtimes$ stands for a tetrahedron and $\epsilon_{\boxtimes}=\pm 1$ for tetrahedra of sublattice A (B). 

The state of a string $|\mathbf{s}_+, \{\vs\}\rangle$ is again parametrized by the location of its $Q=+1$ end $\mathbf s_+$ and by its shape $\{\vs\} \equiv \{\vs_1,\vs_2\ldots\vs_n\}$. String segments $\vs_i$ have four possible orientations: $\vb_0=(1,1,1)/4$, $\vb_1=(-1,1,1)/4$, $\vb_2=(1,-1,1)/4$, and $\vb_3=(-1,-1,1)/4$. A segment with orientation $\vb_0$ or $\vb_3$ must be followed by a segment with orientation $\vb_1$ or $\vb_2$, and vice versa.

The effective Hamiltonian in the subspace of a single string is 
\begin{equation}
H_\mathrm{eff}=-\sqrt{3}J_{z\pm}K_1-J_{\pm}K_2-2J_{\pm\pm}K_3+V
\end{equation}
Kinetic terms $K_1$ and $K_2$ describe first and second-neghbor hopping of the string ends, whereas $K_3$ describes the hopping of a string of length 1. $V = J+nB/\sqrt{3}$ for a string of length $n$. The explicit form of $K_i$ is given in \cite{SP}.

Fig.~\ref{fig:ice-3d} shows the neutron scattering spectrum  $-(\mathrm{Im}S^{aa}+\mathrm{Im}S^{bb})$ calculated with the aid of Lanczos diagonalization, for momentum transfer $\mathbf{k}\parallel\mathbf{B}$ \cite{SP}. We set $J_{\pm}=J_{\pm\pm}=0.36J_{z\pm}$, and $g_{xy}/g_z = 2.4$ as in Yb$_2$Ti$_2$O$_7$ \cite{YbTiO}. The spectral features resemble those of 2D strings (Fig.~\ref{fig:ice-2d}). The branches gradually separate as the string tension increases with $B$. When $J_{z\pm} \sim J_\pm \sim J_{\pm\pm} \ll{}B\ll{}J_{zz}$, the monopole dynamics is dominated by the $x$ and $y$ Zeeman terms whereas the string tension is provided by the $z$ term.

\begin{figure}
\includegraphics[width=0.3\textwidth]{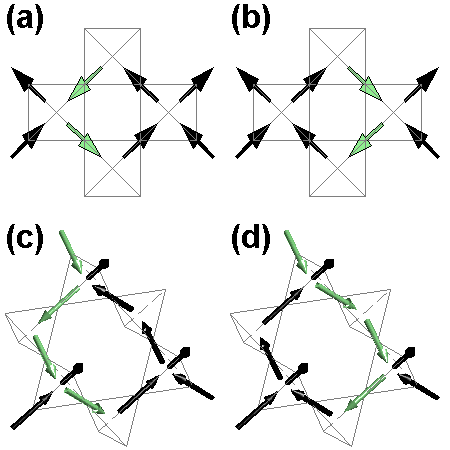}
\caption{Loop-flipping processes in (a,b) checkerboard lattice $|\mathbf{l}\mathbf{r}\rangle\leftrightarrow|\mathbf{r}\mathbf{l}\rangle$ and (c,d) pyrochlore lattice $|\mathbf{b}_2\mathbf{b}_3\mathbf{b}_1\rangle\leftrightarrow|\mathbf{b}_1\mathbf{b}_3\mathbf{b}_2\rangle$.}
\label{fig:shape-fluctuations}
\end{figure}

To the first order in perturbations $J_{z\pm\,,\pm,\,\pm\pm}$, transverse fluctuations induce the motion of a string's ends. At higher-orders in these couplings, the string's shape can change as well. The process involves the simutaneous reversal of spins around a closed loop (minimal length 4 in square ice and 6 in pyrochlore ice) \cite{Barkema, Hermele}. In square ice, a state $|\ldots\mathbf{l}\mathbf{r}\ldots\rangle$ turns into $|\ldots\mathbf{r}\mathbf{l}\ldots\rangle$ and vice versa, Fig.~\ref{fig:shape-fluctuations}. When the position of the monopole and the anti-monopole are both fixed, these fluctuations can be mapped onto a $S=1/2$ XY chain \cite{Kogut}, with spin values $\tau^z = \pm 1/2$ representing $\mathbf r$ and $\mathbf l$ segments, and the Hamiltonian
\begin{equation}
H_\mathrm{fluc} = V_\mathrm{2D}\sum^{n-1}_{i=1}(\tau_i^+\tau_{i+1}^-
	+ \mathrm{H.c.}),
\end{equation}
where $V_\mathrm{2D} = \mathcal O(h^4/J^3)$. Quantum fluctuations reduce tension of the string to $B/\sqrt{2} - 2 |V_\mathrm{2D}|/\pi$. 
When the applied field is below the critical strength $B_c=2\sqrt{2} |V_\mathrm{2D}|/\pi$, the energy cost for string excitations is negative and the fully-polarized state becomes unstable. A similar transition occurs in the pyrochlore quantum spin ice, where a string can be mapped onto an XY chain with second-neighbor interactions only [Fig.~\ref{fig:shape-fluctuations}(c) and (d)],
\begin{equation}
H_\mathrm{fluc}=V_\mathrm{3D}\sum^{n-1}_{i=1}(\tau_{2i-1}^+\tau_{2i+1}^-+\tau_{2i}^+\tau_{2i+2}^- + \mathrm{H.c.}).
\end{equation}
The string tension is reduced by $2|V_{3D}|/\pi$. When $B$ is below the critical value $B_c=2\sqrt{3}|V_\mathrm{3D}|/\pi$, the polarized state becomes unstable.

The fate of the ground state below $B_c$ depends on the dimensionality. On the one hand, the zero field ground state of the pyrochlore spin ice in the perturbative regime $J_{z\pm,\pm,\pm\pm}\ll{}J_{zz}$ is a $U(1)$ spin liquid with deconfined monopoles\cite{Hermele}. Therefore, the transition at $B_c$ could be associated with deconfinement of monopoles. On the other hand, given that the compact quantum electrodynamics is always confined in $2D$ \cite{Polyakov}, the $B=0$ ground state of the 2D quantum spin ice is likely another confined phase separated from the fully-polarized state by the transition at $B_c$.

In the quantum spin-ice material Yb$_2$Ti$_2$O$_7$, the couplings associated with quantum spin fluctuations, viz. $J_{z\pm}$, $J_{\pm}$, and $J_{\pm\pm}$, are comparable with the spin-ice term $J_{zz}$ \cite{YbTiO}. Therefore, perturbative calculations don't apply to it directly. Nonetheless, the physical picture is expected to hold beyond the perturbative regime if the material lies in the phase that is adiabatically connected to the magnetized state. A recent experiment indicates the ground state of Yb$_2$Ti$_2$O$_7$ is a ferromagnet \cite{YbTiO-FM}. The spontaneous magnetization in a $\langle 100 \rangle$ direction acts as a ``molecular field,'' creating nonzero string tension even in the absence of an external field. We expect that strings in quantum spin ice can be detected by neutrons and photons. It would be particularly interesting to observe a continuous evolution of string excitations in an increasing magnetic field applied along a $\langle 100 \rangle$ direction. 

\begin{acknowledgements}
The authors would like to thank Rudro Biswas and Martin Mourigal for useful discussions. Research was supported by the U.S. Department of Energy, Office of Basic Energy Sciences, Division of Materials Sciences and Engineering under Award DE-FG02-08ER46544.
\end{acknowledgements}
\bibliography{string}

\appendix*

\begin{widetext}
\begin{center}
\textbf{\large Supplementary Material}
\end{center}
\renewcommand{\thefigure}{S\arabic{figure}}
\renewcommand{\theequation}{S\arabic{equation}}
\setcounter{equation}{0}
\setcounter{figure}{0}

\section{Even and odd parity states of quantum strings in 2D}
We define the following reflection operators $X_c$ (Fig.2):
\begin{equation}
X_c|\mathbf{s}_+,\{\mathbf{s}_1\dots\mathbf{s}_m\dots\mathbf{s}_n\}\rangle=\left\{
\begin{array}{ll}
|\mathbf{s}_+,\{\mathbf{s}_1\dots\bar{\mathbf{s}}_m\dots\mathbf{s}_n\}\rangle; & m=c-c_++1\\
|\mathbf{s}_+,\{\mathbf{s}_1\dots\mathbf{s}_m\dots\mathbf{s}_n\}\rangle; & c<c_+\;\mathrm{or}\;{}c>c_++n-1
\end{array}
\right.
\end{equation}
Here we have introduced shorthand notation $\bar{\mathbf{l}}=\mathbf{r}$ and $\bar{\mathbf{r}}=\mathbf{l}$, and $c_+$ is the $c$ component of the monopole position vector $\mathbf{s}_+$. $X_c$ reflects the orientation of the segment located at $c$. If no segment is located at $c$, $X_c$ effectively acts on vacuum, and therefore we define that $X_c$ acts trivially on such a state. It can be seen that $X^2_c=1$ and $[X_c,X_{c'}]=0$. 

Now we consider the subspace of $k_b=0$. The effect of $X_c$ acting on $|k_b=0,c_+,\{\mathbf{s}_i\}\rangle$ is given by:
\begin{equation}
X_c|k_b=0,c_+,\{\mathbf{s}_1\dots\mathbf{s}_m\dots\mathbf{s}_n\}\rangle=\left\{
\begin{array}{ll}
|k_b=0,c_+,\{\mathbf{s}_1\dots\bar{\mathbf{s}}_m\dots\mathbf{s}_n\}\rangle; & m=c-c_++1\\
|k_b=0,c_+,\{\mathbf{s}_1\dots\mathbf{s}_m\dots\mathbf{s}_n\}\rangle; & c<c_+\;\mathrm{or}\;{}c>c_++n-1
\end{array}
\right.
\end{equation}
Hence, $k_b=0$ states form an invariant subspace of $X_c$ operators. When $k_b=0$, the effective Hamiltonian (4) becomes:
\begin{align}
H_\mathrm{eff}|c_{+},\{\mathbf{s}_1\ldots \mathbf{s}_{n}\}\rangle 
&= (J + nB/\sqrt{2})|c_{+},\{\mathbf{s}_1\ldots\mathbf{s}_{n}\}\rangle 
-(h/2)|c_{+},\{\mathbf{s}_1\ldots\mathbf{s}_{n-1}\}\rangle
-(h/2)|c_{+}+1,\{\mathbf{s}_2\ldots\mathbf{s}_{n}\}\rangle
\nonumber\\	
&  - \ (h/2) \sum_{\mathbf{s}_{n+1}}|c_{+},\{\mathbf{s}_1\ldots\mathbf{s}_{n+1}\}\rangle
- \ (h/2) \sum_{\mathbf{s}_{0}}|c_{+}-1,\{\mathbf{s}_0\ldots\mathbf{s}_{n}\}\rangle.\label{blockH}
\end{align}
We have dropped the $k_b$ index for clarity. It can be seen that $[X_c,H_\mathrm{eff}]=0$ in the $k_b=0$ subspace. In what follows, we present eigenstates and eigenvalues of the above Hamiltonian in different parity sectors.

\subsection{All-even parity states}
We consider the all-even states for which $X_c=1$ for all $c$:
\begin{equation}
|c_+,n\rangle=2^{-n/2}\sum_{\mathbf{s}_1\dots\mathbf{s}_n}|c_+,\mathbf{s}_1\dots\mathbf{s}_n\rangle
\end{equation}
Acting $H_\mathrm{eff}$ on $|c_+,n\rangle$,
\begin{equation}
H_\mathrm{eff}|c_+,n\rangle=(J+\frac{nB}{\sqrt{2}})|c_+,n\rangle-\frac{h}{\sqrt{2}}|c_+,n+1\rangle-\frac{h}{\sqrt{2}}|c_+,n-1\rangle-\frac{h}{\sqrt{2}}|c_+-1,n+1\rangle-\frac{h}{\sqrt{2}}|c_++1,n-1\rangle\label{even}
\end{equation}
The effective Hamiltonian (\ref{even}) is reduced to a one dimensional two-body problem where one particle is located at $c_+$ and the other is at $c_-=c_++n>c_+$ (Fig.\ref{fig1}). Note that (\ref{even}) is translationally invariant in $c$, and the states with momentum $k_c$ are henceforth constructed (6):
\begin{equation}
|k_c,n\rangle=\sum_{c_+}e^{ik_c(c_++c_-)/2}|c_+,n\rangle=2^{-n/2}\sum_{c_+}\sum_{\mathbf{s}_1\dots\mathbf{s}_n}e^{ik_c(c_++c_-)/2}|c_+,\mathbf{s}_1\dots\mathbf{s}_n\rangle\label{alleven}
\end{equation}
$H_\mathrm{eff}$ becomes block-diagonalized as shown in Eq.(7):
\begin{equation}
H_\mathrm{eff}|k_c,n\rangle=(J+\frac{nB}{\sqrt{2}})|k_c,n\rangle-\sqrt{2}h\cos(\frac{k_c}{2})|k_c,n+1\rangle-\sqrt{2}h\cos(\frac{k_c}{2})|k_c,n-1\rangle
\end{equation}
We shall not repeat the solution in the all-even sector, which has been presented in the main text.

\begin{figure}
\centering
\includegraphics[width=0.5\textwidth]{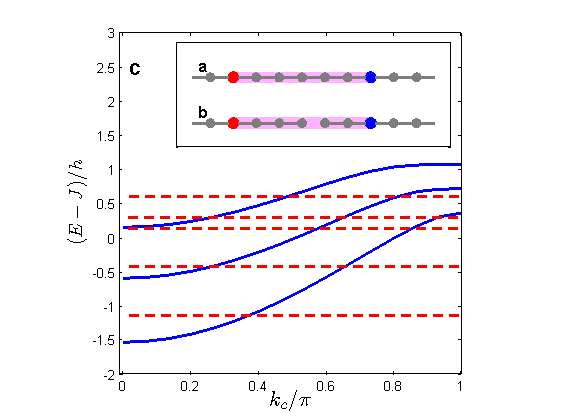}
\caption{(a) The effective one-dimensional problem for the all-even sector. The string (purple rectangle) binds a monopole (red dot) and an anti-monopole (blue dot). (b) The effective one-dimensional problem for a odd-parity sector. (c) The dispersion of the first few even (blue solid line) and odd (red dashed line) strings modes. $B/h=0.5$.}\label{fig1}
\end{figure}

\subsection{Odd-parity states}
Now we discuss odd-parity sectors. We consider states that are odd under one $X_c$ operator and even under all the others. It doesn't exhaust all possibilities, yet our discussion on this special case can be generalized to more complicated situations straightforwardly.

Without loss of generality, we impose the condition: $X_0=-1$, and $X_c=1$ for $c\neq0$:
\begin{equation}
|c_+,n\rangle=2^{-n/2}\sum_{\mathbf{s}_1\dots{}\mathbf{s}_n}(-)^{\chi(1-c_+)}|c_+,\mathbf{s}_1\dots\mathbf{s}_n\rangle\quad{}1-n\le{}c_+\le{}0
\end{equation}
$\chi(m)=1(-1)$ for $\mathbf{s}_m=\mathbf{l}(\mathbf{r})$. We mark such a sector as $\{\mathrm{odd},c=0\}$ where $c$ stands for the position where the odd-parity condition is imposed. Acting $H_\mathrm{eff}$ on $|c_+,n\rangle$, we yield
\begin{subequations}\label{odd}
\begin{align}
&H_\mathrm{eff}|c_+,n\rangle=(J+nB/\sqrt{2})|c_+,n\rangle-(h/\sqrt{2})(\sum_{m=n\pm1}|c_+,m\rangle+|c_+-1,n+1\rangle+|c_++1,n-1\rangle) \; (1-n<c_+<0)\\
&H_\mathrm{eff}|0,n\rangle=(J+nB/\sqrt{2})|0,n\rangle-(h/\sqrt{2})(\sum_{m=n\pm1}|0,m\rangle+|-1,n+1\rangle) \; (n\ge2)\\
&H_\mathrm{eff}|1-n,n\rangle=(J+nB/\sqrt{2})|1-n,n\rangle-(h/\sqrt{2})(|1-n,n+1\rangle+|-n,n+1\rangle+|2-n,n-1\rangle) \; (n\ge2)\\
&H_\mathrm{eff}|0,1\rangle=(J+B/\sqrt{2})|0,1\rangle-(h/\sqrt{2})(|0,2\rangle+|-1,2\rangle)
\end{align}
\end{subequations}
Once again, Hamiltonian (\ref{odd}) is equivalent to a one dimensional two-body problem in which one particle is located at $c_+$ and the other at $c_-=c_++n>c_+$. We see that the odd parity condition at $c=0$ effectively impose a constraint on the particle coordinates; the monopole is constrained to $c_+\le0$ whereas the anti-monopole $c_-\ge1$ (Fig.\ref{fig1}). The monopole and the anti-monopole are localized near $c=0$ due to the binding force between them. Therefore, in terms of string, the odd-parity condition imposed at $c$ effectively pins the string at the very same location. This is in contrast to the all-even sector where strings are free to move.

In the limit of $B\ll{}h$, the continuum approximation gives the energy spectrum
\begin{equation}
E_{i,j}=J-2\sqrt{2}h-B/\sqrt{2}+(\lambda_i+\lambda_j)(hB^2)^\frac{1}{3}/\sqrt{2}
\end{equation}
where $\lambda_i$ are zeros of Airy function. Different from eigenstates in the all-even sector, the eigenstates in the odd sector $\{odd,c\}$ are labeled by two indices $i,j$, each corresponding to one end of the string. In the limit of $B\gg{}h$, perturbation theory gives the energy of the lowest eigenstate
\begin{equation}
E_{0,0}=J+B/\sqrt{2}-\sqrt{2}h^2/B
\end{equation}

The eigenstates in two different sectors $\{\mathrm{odd},c\}$ and $\{\mathrm{odd},c'\}$ are related through rigid translation. Therefore, we can construct the Bloch states with crystal momentum $k_c$:
\begin{equation}
|k_c,\{i,j\}\rangle=\sum_{r}e^{ik_cr}|\{i,j\},r\rangle
\end{equation}
where $|\{i,j\},r\rangle$ is the eigenstate $\{i,j\}$ in sector $\{\mathrm{odd},r\}$. Since the odd-parity states are all localized, the band dispersion of the Bloch state is flat:
\begin{equation}
E_{i,j}(k_c)=E_{i,j}
\end{equation}
in contrast to the dispersive bands of all-even states.

\section{The explicit form of $K_{i}$}
We introduce the momentum basis for strings in pyrochlore quantum spin ice:
\begin{equation}
|\mathbf{k},\{\mathbf{s}_i\}\rangle=\sum_{\mathbf{s}_+}e^{i\mathbf{k}\cdot(\mathbf{s}_++\mathbf{s}_-)/2}|\mathbf{s}_+,\{\mathbf{s}_i\}\rangle.
\end{equation}
Here $\mathbf{s}_+$ and $\mathbf{s}_i$ have been defined in the main text. For a string with momentum $\mathbf k$,
\begin{subequations}
\begin{align}
K_1|\{\mathbf{s}_1\ldots\mathbf{s}_n\}\rangle
	&= \gamma^\ast(\mathbf{s}_{n-1},\mathbf{s}_n)e^{i\mathbf{k}\cdot{}\mathbf{s}_n/2}|\{\mathbf{s}_1\ldots{}\mathbf{s}_{n-1}\}\rangle
	+ \sum_{\mathbf{s}_{n+1}}\gamma(\mathbf{s}_n,\mathbf{s}_{n+1})e^{-i\mathbf{k}\cdot\mathbf{s}_{n+1}/2}|\{\mathbf{s}_1\ldots\mathbf{s}_{n+1}\}\rangle\nonumber\\
	&+\gamma^\ast(\mathbf{s}_2,\mathbf{s}_1)e^{-i\mathbf{k}\cdot\mathbf{s}_1/2}|\{\mathbf{s}_2\ldots\mathbf{s}_n\}\rangle
	+\sum_{\mathbf{s}_0}\gamma(\mathbf{s}_1,\mathbf{s}_0)e^{i\mathbf{k}\cdot\mathbf{s}_0/2}|\{\mathbf{s}_0\ldots\mathbf{s}_n\}\rangle\\
K_2|\{\mathbf{s}_1\ldots\mathbf{s}_n\}\rangle
	&= e^{i\mathbf{k}\cdot(\mathbf{s}_{n-1}+\mathbf{s}_n)/2}|\{\mathbf{s}_1\ldots\mathbf{s}_{n-2}\}\rangle
	+ e^{-i\mathbf{k}\cdot(\bar{\mathbf{s}}_{n}-\mathbf{s}_n)}|\{\mathbf{s}_1\ldots{}\bar{\mathbf{s}}_{n}\}\rangle
	+ \sum_{\mathbf{s}_{n+1},\mathbf{s}_{n+2}}e^{-i\mathbf{k}\cdot(\mathbf{s}_{n+1}+\mathbf{s}_{n+2})/2}|\{\mathbf{s}_1\ldots\mathbf{s}_{n+2}\}\rangle
\nonumber\\
	&+e^{-i\mathbf{k}\cdot(\mathbf{s}_1+\mathbf{s}_2)/2}|\{\mathbf{s}_3\ldots\mathbf{s}_n\}\rangle
	+e^{i\mathbf{k}\cdot(\bar{\mathbf{s}}_1-\mathbf{s}_1)}|\{\bar{\mathbf{s}}_1\ldots\mathbf{s}_{n}\}\rangle
	+\sum_{\mathbf{s}_{-1},\mathbf{s}_0}e^{i\mathbf{k}\cdot(\mathbf{s}_{-1}+\mathbf{s}_0)/2}|\{\mathbf{s}_{-1}\ldots\mathbf{s}_n\}\rangle\\
K_3|\mathbf{b}_{0,3}\rangle&=
	\sum_{i=1,2}\zeta^\ast_{[0,3]i}
	\cos{\left(\mathbf{k}\cdot\frac{\mathbf{b}_{0,3}+\mathbf{b}_i}{2}\right)}|\mathbf{b}_i\rangle;\quad{}K_3|\mathbf{b}_{1,2}\rangle=
	\sum_{i=0,3}\zeta_{[1,2]i}
	\cos{\left(\mathbf{k}\cdot\frac{\mathbf{b}_{1,2}+\mathbf{b}_i}{2}\right)}|\mathbf{b}_i\rangle.
\end{align}
\end{subequations}
We have omitted momentum index $\mathbf{k}$ for clarity. Here $\gamma(\mathbf{b}_i,\mathbf{b}_j)\equiv\gamma_{ij}$ is a $4\times4$ anti-Hermitian matrix with nonzero elements $\gamma_{01}=i - r e^{-i\pi/3}$, $\gamma_{02}=e^{i5\pi/6} - r e^{-i\pi/3}$, $\gamma_{13}= e^{i\pi/6} + r e^{i\pi/3}$, $\gamma_{23} = i + r e^{i\pi/3}$; and $r = g_{xy}B/(3\sqrt{2}g_{z}J_{z\pm})$. The short-hand notation $\bar{\mathbf s}_i$ means exchanging $\mathbf{b}_0 \leftrightarrow \mathbf{b}_3$ and $\mathbf{b}_1 \leftrightarrow \mathbf{b}_2$.

\section{Calculation of the dynamical structure factors}
In this section, we describe the technical details of calculating the dynamical structure factors $S^{\alpha\alpha}(\mathbf{k},\omega)$,  $\alpha=a,\,b,\,c$. The imaginary part of the dynamical structure factor is given by Kubo formula,
\begin{equation}
-\mathrm{Im}S^{\alpha\alpha}(\mathbf{k},\omega)/\pi=\sum_{n}|f^\alpha_n(\mathbf{k})|^2\delta(\omega-E_n+E_G).
\end{equation}
The summation is over all excited states $n$ with energy $E_n$. The scattering amplitude $f^\alpha_n(\mathbf{k})$ is given by
\begin{equation}
f^\alpha_n(\mathbf{k})=\sum_\mathbf{R}e^{i\mathbf{k}\cdot\mathbf{R}}\langle{}n|S^\alpha_{\mathbf{R}}|G\rangle
\end{equation}
Here $|G\rangle$ is the ground state. $\mathbf{R}$ is the spatial position of operator $S^\alpha_\mathbf{R}$.

We discuss the 2D quantum spin ice at first. Here we just consider $S^{aa}$ for simplicity. Within the leading order in $h/J$, the ground state is the fully-polarized state. The effect of $S^{c}_{\mathbf{R}}$ on the ground state is to flip the spin at $\mathbf{R}$ and equivalently to create a string of length 1. Therefore,
\begin{equation}
\sum_\mathbf{R}S^a_{\mathbf{R}}e^{i\mathbf{k}\cdot\mathbf{R}}|G\rangle=\frac{1}{\sqrt{2}}(e^{i\mathbf{k}\cdot\mathbf{l}/2}|\mathbf{k},\mathbf{l}\rangle+e^{i\mathbf{k}\cdot\mathbf{r}/2}|\mathbf{k},\mathbf{r}\rangle)\equiv|\mathbf{k},\psi\rangle
\end{equation}
up to a normalization constant. Here $|\mathbf{k},\mathbf{l}/\mathbf{r}\rangle$ is the length 1 string state with total momentum $\mathbf{k}$ and bond orientation $\mathbf{l}/\mathbf{r}$. The low energy scattering is dominated by the scattering processes between the ground state and single string states. Therefore, we sum over all single-string states instead of the whole Hilbert space. The scattering amplitude is determined by the overlap between $|\mathbf{k},\psi\rangle$ and the single-string eigenstate $|\mathbf{k},n\rangle$ of $H_\mathrm{eff}$:
\begin{equation}
f_n(\mathbf{k})=\langle{}\mathbf{k},n|\mathbf{k},\psi\rangle
\end{equation}
We obtain the formula for calculating $-\mathrm{Im}S^{aa}(\mathbf{k},\omega)$ in practice:
\begin{equation}
-\frac{1}{\pi}\mathrm{Im}S^{aa}(\mathbf{k},\omega)=-\frac{1}{\pi}\mathrm{Im}\langle\mathbf{k},\psi|\frac{1}{\omega+i0^+-H_\mathrm{eff}}|\mathbf{k},\psi\rangle.\label{dyna2D}
\end{equation}
The standard Lanczos technique is used to evaluate (\ref{dyna2D}). Note that $H_\mathrm{eff}$ is block-diagonalized in $\mathbf{k}$ and only the states with momentum $\mathbf{k}$ are needed. For each $\mathbf{k}$, $\Lambda_\mathrm{max}=100\sim150$ Lanczos iterations are used to construct Lanczos basis vectors, starting from the initial state $|\mathbf{k},\psi\rangle$. The string states are truncated at maximum length $L_\mathrm{max}=15\sim20$ due to the limitation of computer memory size. The convergence is checked by varying $\Lambda_\mathrm{max}$ and $L_{\mathrm{max}}$. A finite but small broadening factor $\tau$ is used to replace the infinitesimal positive number $0^+$ in (\ref{dyna2D}). 

For the special case $k_b=0$, the calculation of (\ref{dyna2D}) is greatly simplified. The initial state $|\mathbf{k},\psi\rangle$ becomes
\begin{equation}
|\mathbf{k},\psi\rangle=\frac{1}{\sqrt{2}}(|\mathbf{k},\mathbf{l}\rangle+|\mathbf{k},\mathbf{r}\rangle)=|k_c,n=1\rangle_\mathrm{even}
\end{equation}
which is an all-even state. Therefore only even-parity states $|k_c,n\rangle$ (\ref{alleven}) are needed to calculate $S^{aa}(k_b=0,\omega)$.

The calculation of dynamical structure factors of pyrochlore quantum spin ice is carried out in the same manner. The corresponding formula is
\begin{equation}
-\frac{1}{\pi}\mathrm{Im}S^{\alpha\alpha}(\mathbf{k},\omega)=-\frac{1}{\pi}\mathrm{Im}\langle\mathbf{k},\psi_\alpha|\frac{1}{\omega+i0^+-H_\mathrm{eff}}|\mathbf{k},\psi_\alpha\rangle\label{dyna3D}
\end{equation}
where $H_\mathrm{eff}$ is the effective Hamiltonian for a single string in pyrochlore quantum spin ice. When the momentum transfer $\mathbf{k}\parallel{}c$, only the sum $S^{aa}+S^{bb}$ contributes to the total scattering intensity due to the transverse polarization factor of neutrons. The relevant initial states are
\begin{subequations}
\begin{align}
|\psi_a\rangle&=\frac{1}{2}(e^{i\mathbf{k}\cdot\mathbf{R}_0}|\mathbf{k},\mathbf{b}_0\rangle+e^{i\mathbf{k}\cdot\mathbf{R}_1}|\mathbf{k},\mathbf{b}_1\rangle-e^{i\mathbf{k}\cdot\mathbf{R}_2}|\mathbf{k},\mathbf{b}_2\rangle-e^{i\mathbf{k}\cdot\mathbf{R}_3}|\mathbf{k},\mathbf{b}_3\rangle)\\
|\psi_b\rangle&=\frac{1}{2}(e^{i(\mathbf{k}\cdot\mathbf{R}_0+\pi/3)}|\mathbf{k},\mathbf{b}_0\rangle-e^{i(\mathbf{k}\cdot\mathbf{R}_1-\pi/3)}|\mathbf{k},\mathbf{b}_1\rangle+e^{i(\mathbf{k}\cdot\mathbf{R}_2-\pi/3)}|\mathbf{k},\mathbf{b}_2\rangle-e^{i(\mathbf{k}\cdot\mathbf{R}_3+\pi/3)}|\mathbf{k},\mathbf{b}_3\rangle)
\end{align}
\end{subequations}
for $S^{aa}$ and $S^{bb}$ respectively. Here $\mathbf{R}_i$ are spatial coordinates of the four inequivalent sites in pyrochlore lattice. The phase factors are due to the mismatch between the local spin frames $(x,y,z)$ and global spin frame $(a,b,c)$.
\end{widetext}
\end{document}